\documentclass[fleqn,usenatbib]{mn2e}

\usepackage{graphicx}	
\usepackage{amssymb}
\usepackage{amsmath}
\usepackage{hyperref}

\def\jnl@style{}
\def\aaref@jnl#1{{\jnl@style#1}}

\def\aaref@jnl#1{{\jnl@style#1}}

\def\aj{\aaref@jnl{AJ}}                   
\def\araa{\aaref@jnl{ARA\&A}}             
\def\apj{\aaref@jnl{ApJ}}                 
\def\apjl{\aaref@jnl{ApJ}}                
\def\apjs{\aaref@jnl{ApJS}}               
\def\ao{\aaref@jnl{Appl.~Opt.}}           
\def\apss{\aaref@jnl{Ap\&SS}}             
\def\aap{\aaref@jnl{A\&A}}                
\def\aapr{\aaref@jnl{A\&A~Rev.}}          
\def\aaps{\aaref@jnl{A\&AS}}              
\def\azh{\aaref@jnl{AZh}}                 
\def\baas{\aaref@jnl{BAAS}}               
\def\jrasc{\aaref@jnl{JRASC}}             
\def\memras{\aaref@jnl{MmRAS}}            
\def\mnras{\aaref@jnl{MNRAS}}             
\def\pra{\aaref@jnl{Phys.~Rev.~A}}        
\def\prb{\aaref@jnl{Phys.~Rev.~B}}        
\def\prc{\aaref@jnl{Phys.~Rev.~C}}        
\def\prd{\aaref@jnl{Phys.~Rev.~D}}        
\def\pre{\aaref@jnl{Phys.~Rev.~E}}        
\def\prl{\aaref@jnl{Phys.~Rev.~Lett.}}    
\def\pasp{\aaref@jnl{PASP}}               
\def\pasj{\aaref@jnl{PASJ}}               
\def\qjras{\aaref@jnl{QJRAS}}             
\def\skytel{\aaref@jnl{S\&T}}             
\def\solphys{\aaref@jnl{Sol.~Phys.}}      
\def\sovast{\aaref@jnl{Soviet~Ast.}}      
\def\ssr{\aaref@jnl{Space~Sci.~Rev.}}     
\def\zap{\aaref@jnl{ZAp}}                 
\def\nat{\aaref@jnl{Nature}}              
\def\iaucirc{\aaref@jnl{IAU~Circ.}}       
\def\aplett{\aaref@jnl{Astrophys.~Lett.}} 
\def\apspr{\aaref@jnl{Astrophys.~Space~Phys.~Res.}}
\def\bain{\aaref@jnl{Bull.~Astron.~Inst.~Netherlands}} 
\def\fcp{\aaref@jnl{Fund.~Cosmic~Phys.}}  
\def\gca{\aaref@jnl{Geochim.~Cosmochim.~Acta}}   
\def\grl{\aaref@jnl{Geophys.~Res.~Lett.}} 
\def\jcp{\aaref@jnl{J.~Chem.~Phys.}}      
\def\jgr{\aaref@jnl{J.~Geophys.~Res.}}    
\def\jqsrt{\aaref@jnl{J.~Quant.~Spec.~Radiat.~Transf.}}
\def\memsai{\aaref@jnl{Mem.~Soc.~Astron.~Italiana}}
\def\nphysa{\aaref@jnl{Nucl.~Phys.~A}}   
\def\physrep{\aaref@jnl{Phys.~Rep.}}   
\def\physscr{\aaref@jnl{Phys.~Scr}}   
\def\planss{\aaref@jnl{Planet.~Space~Sci.}}   
\def\procspie{\aaref@jnl{Proc.~SPIE}}   

\begin{document}

\title[Resolving the Planetesimal Belt of HR~8799 with ALMA]{Resolving the Planetesimal Belt of HR~8799 with ALMA}
\author[M. Booth et al.]{Mark Booth$^{1,2}$\thanks{E-mail: markbooth@cantab.net}, Andr\'es Jord\'an$^{1,3}$, Simon Casassus$^{2,4}$, Antonio S. Hales$^{5,6}$, \newauthor William R. F. Dent$^{5}$, Virginie Faramaz$^{1}$, Luca Matr\`a$^{7,8}$, Denis Barkats$^{9}$, \newauthor Rafael Brahm$^{1,3}$ and Jorge Cuadra$^{1,2}$  \\
$^{1}$ Instituto de Astrof\'isica, Pontificia Universidad Cat\'olica de Chile, Vicu\~na Mackenna 4860, Santiago, Chile \\
$^{2}$ Millennium Nucleus ``Protoplanetary Disks'' \\
$^{3}$ Millennium Institute of Astrophysics, Vicu\~na Mackenna 4860, Santiago, Chile\\
$^{4}$ Departamento de Astronomia, Universidad de Chile, Casilla 36-D, Santiago, Chile \\
$^{5}$ Joint ALMA Observatory, Alonso de C\'ordova 3107, Vitacura 763-0355, Santiago, Chile \\
$^{6}$ National Radio Astronomy Observatory, 520 Edgemont Road, Charlottesville, Virginia, 22903-2475, USA \\
$^{7}$ Institute of Astronomy, University of Cambridge, Madingley Road, Cambridge CB3 0HA, UK \\
$^{8}$ European Southern Observatory, Alonso de C\'ordova 3107, Vitacura, Casilla 19001, Santiago, Chile \\
$^{9}$ Harvard University, 60 Garden Street, Cambridge, MA 02138, USA 
}

\date{Accepted 2016 March 15. Received 2016 March 14; in original form 2015 December 21}
\pubyear{2016}

\maketitle

\begin{abstract}
The star HR 8799 hosts one of the largest known debris discs and at least four giant
planets. Previous observations have found evidence for a warm belt within the orbits
of the planets, a cold planetesimal belt beyond their orbits and a halo of small grains. With the infrared data, it is
hard to distinguish the planetesimal belt emission from that of the grains in the halo. With this in mind, the
system has been observed with ALMA in band 6 (1.34 mm) using a compact array format. These observations allow the inner edge of the planetesimal belt to be resolved
for the first time. A radial distribution of dust grains is fitted to the data using an MCMC method. The disc is best fit by a broad ring between $145^{+12}_{-12}$~AU and $429^{+37}_{-32}$~AU at an inclination of $40^{+5}_{-6}{{\degr}}$ and a position angle of $51^{+8}_{-8}{{\degr}}$. A disc edge at $\sim$145~AU is too far out to be explained simply
by interactions with planet b, requiring either a more complicated dynamical history
or an extra planet beyond the orbit of planet b. 

\end{abstract}


\section{Introduction}
As we build up our census of nearby stars, we find many systems that are host to both a debris disc and planets. These systems provide interesting test cases for understanding how planets and discs interact, the most useful of which are the cases where the disc has been resolved, allowing us to determine the geometry of the system and look for asymmetries in the disc \citep[see][for a review]{moro13}. Unfortunately, resolved images to date have only  probed  the outer reaches of planetary systems (the Kuiper belt analogues), whereas the majority of detected planets (detected through radial velocity and transit observations) are very close to their host stars. This discrepancy in the measured scales of detection of host planets and host disc has limited our understanding of their interaction. The advent of direct imaging of planets is changing this paradigm as this method is much more sensitive to planets at tens of AU. So far HR~8799 is the only star around which multiple planets have been detected through direct imaging \citep{marois08,marois10}, making it a key system for detailed investigations.

Observations of the debris disc around HR~8799 go back to IRAS \citep{sadakane86}. Detailed study of this disc did not happen until it was observed by Spitzer \citep{su09}. This resulted in a detailed SED but also a resolved image at 24~$\mu$m. The Spitzer observations imply that the dust must be going out to very large radii. This lead \citet{su09} to propose a model of the debris disc that consists of an inner asteroid belt analogue, a planetesimal belt between 100 and 310~AU and a blowout grain halo going out to at least 1500~AU. Deep observations using Herschel are also fit well by this model and have a high enough resolution that it is possible to constrain the inclination of the system to 26$\pm3{\degr}$ \citep{matthews14}. This agrees with some measures of the inclination of the star \citep{reidemeister09} and the likely inclination of the planets' orbits \citep{gozdziewski14}. The disc has previously been observed in the sub-mm using the James Clerk Maxwell Telescope \citep{williams06}, Caltech Submillimeter Observatory \citep{patience11} and the Submillimeter Array \citep{hughes11}, although none of these had the resolution and/or sensitivity required to accurately determine the geometry of the cold belt.

\section{Observations}
\label{sobs}
\begin{figure*}
	\centering
	\includegraphics[height=5.9cm]{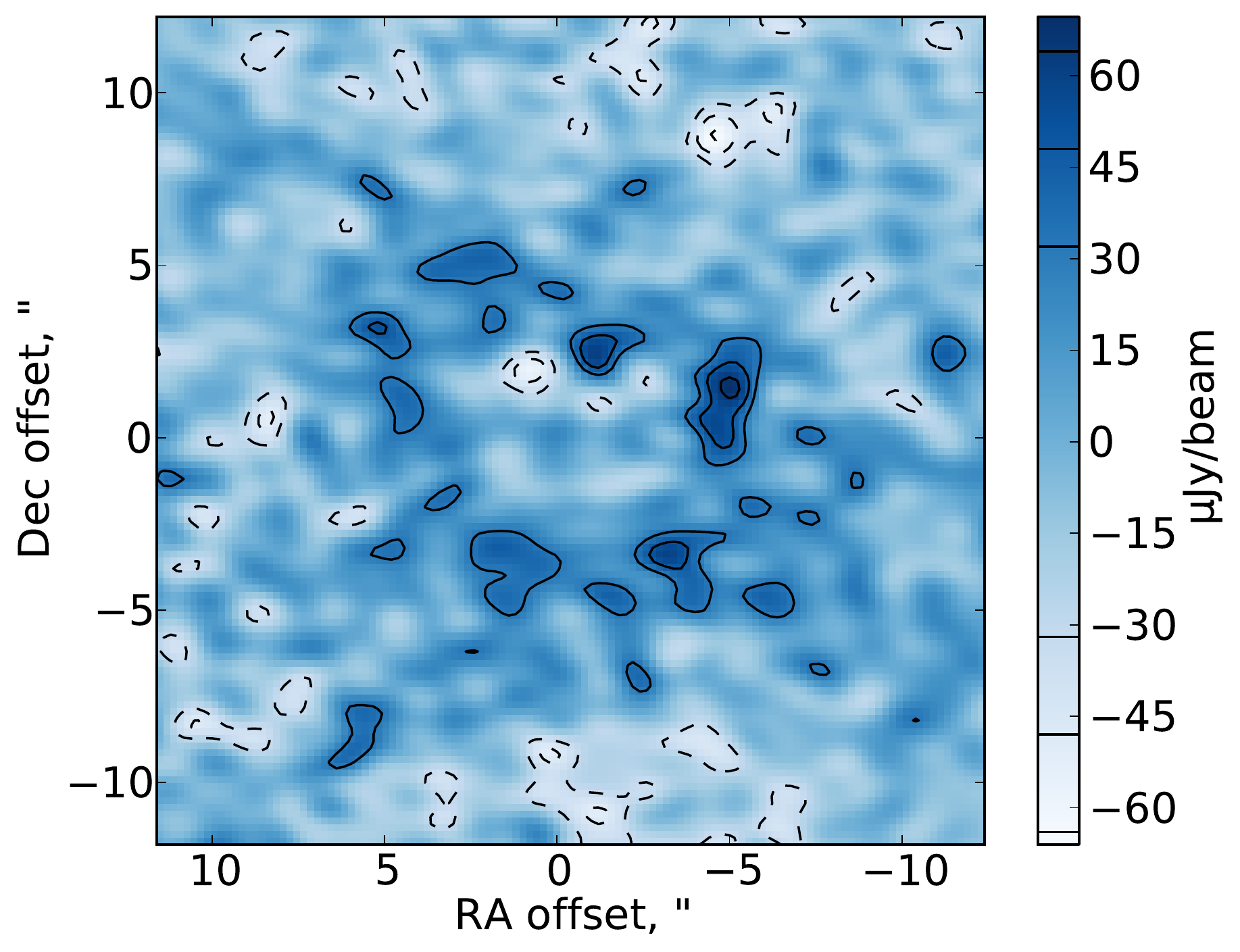}
	\hspace{1cm}
	\includegraphics[height=5.9cm]{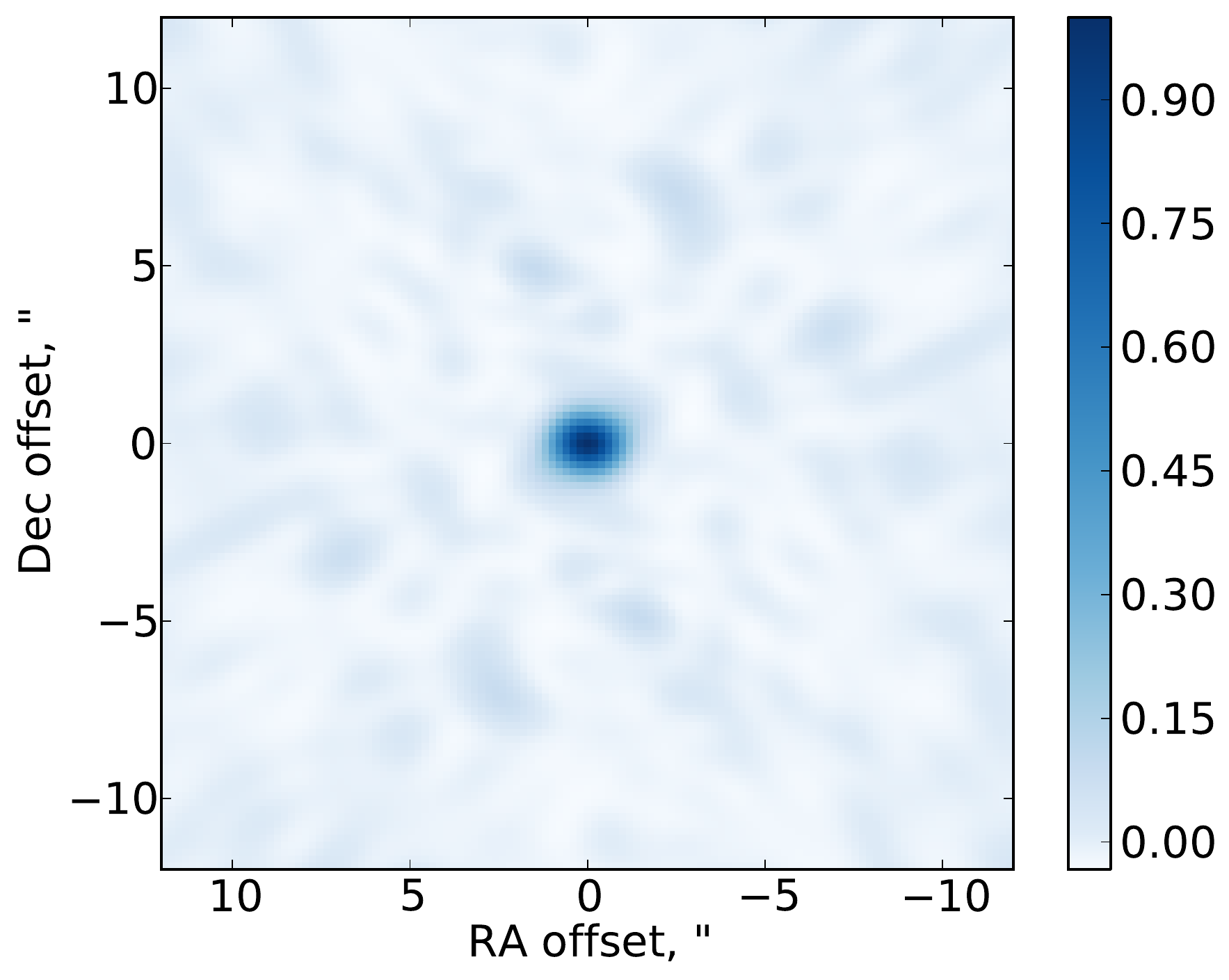}
	\caption{Dirty image (left) of the continuum emission and the dirty beam (right). The dirty image is the inverse Fourier transform of the observed complex visibilities and has been created using natural weighting and multi-frequency synthesis. Contours show $\pm2$, 3 and 4-$\sigma$ significance levels. Negative `emission' is seen surrounding the ring, which is due to the sidelobes of the beam that need to be removed by the modelling process (see section \ref{smod}).}
	\label{fclean}
\end{figure*}
HR 8799 was observed with ALMA in band 6 (1.34~mm) for the cycle 1 project 2012.1.00482.S. The observations were taken with a compact array format with baselines between 15-349~m. Given the distance to the star of 39.4~pc \citep{leeuwen07}, these correspond to spatial scales of 32 and 430~AU respectively. The data are comprised of 5 separate observations taken between 3rd and 18th January 2015. Details are shown in table \ref{tobs}. The correlator was configured to optimise continuum sensitivity,
processing two polarizations in four, 2 GHz-wide basebands centred at 216, 218, 231 and 233 GHz,
each with 128 spectral channels of width 16~MHz (20~kms$^{-1}$).
\begin{table}%
\begin{tabular}{cccc}
ID & Observation Start & Antennas & PWV (mm) \\ 
\hline
1 & 2015-01-03 20:23:49 & 38 & 1.2 \\ 
2 & 2015-01-13 22:03:07 & 37 & 3.6 \\ 
3 & 2015-01-14 21:47:52 & 38 & 4.2 \\ 
4 & 2015-01-16 21:44:55 & 35 & 3.5 \\ 
5 & 2015-01-18 18:35:04 & 34 & 4.0 \\ 
\end{tabular}
\caption{Details of the observations. PWV stands for precipitable water vapour.}
\label{tobs}
\end{table}

The data was calibrated using standard observatory calibration in \texttt{CASA} version 4.3.1, which includes water vapour radiometry correction, system temperature, complex gain calibration and flagging. Uranus was used as a flux calibrator for observations 1-4 and Neptune was used for observation 5. J2253+1608 was used to calibrate the phase. Calibration of observation 2 was worse than the rest due to large water vapour fluctuations so this data was discarded from the analysis. The total on source time is 2.8 hours. Figure \ref{fclean} shows the dirty image (the inverse Fourier transform of the observed visibilities) of the continuum emission along with the dirty beam. These were created using \texttt{CLEAN} with zero iterations and a natural weighting, taking advantage of the multi-frequency synthesis. The resulting synthesised beam has a size 1.7\arcsec$\times$1.3{\arcsec}. The RMS noise is best measured using a large region far from the source so as not to be affected by any of the artefacts of the Fourier transform. In this case, the diameter of the source (see section \ref{smod}) is comparable to the FWHM of the primary beam (26\arcsec), therefore the RMS was measured by creating a dirty image pointing slightly away from the target (+90{\arcsec} in declination) with a field of view that does not include the target. Using this method, the RMS is measured to be $\sigma=16$~$\mu$Jy/beam. A ring is detected at $\sim$5.5{\arcsec}. This ring is surrounded by negative emission and an outer `ring' is also present at $\sim$20{\arcsec}, both of which are artefacts of the dirty beam.

As well as continuum, we detected CO J=2-1 emission at a barycentric velocity consistent with that of the star (-12.6$\pm$1.4 km/s). The emission is spectrally unresolved at our resolution of 40.6 km/s, in line with previous CO measurements showing much narrower line widths \citep{williams06,su09,hughes11}. Spatially, the dirty map shows complicated structure, with striping along the NE-SW direction analogous to that reported from previous SMA observations and indicative of line-of-sight contamination by the molecular cloud HLCG 92-35 \citep{yamamoto03}. The CO emission is seen as a clump $\sim$12" SW of the star (near the continuum ansa), which appears to be consistent with that detected in previous JCMT observations \citep{williams06,su09}. Our ALMA observations therefore confirm the presence of CO emission along the line of sight to HR8799 originating from the HLCG 92-35 cloud.

\section{Modelling}
\label{smod}

Interferometric observations measure complex visibilities in the $u$-$v$ plane -- the Fourier transform of the image plane. Whilst it is common practice to model interferometric observations in the Fourier plane, it is more efficient, whilst still retaining accuracy, to work in the image plane as the dirty image is simply the Fourier transform of the gridded visibilities \citep[see][for the basics of interferometry]{clark99}.

To find the best fit parameters of the disc and their uncertainties an MCMC routine is run making use of \texttt{emcee} \citep{foreman13}. The disc is modelled with 6 free parameters using a line of sight integrator method \citep{wyatt99}. We assume a wide ring of dust grains between $R_{in}$ and $R_{out}$ with an optical depth that varies as $R^\gamma$. The ring is inclined from face on by an angle $I$ and given a position angle, $\Omega$, measured anti-clockwise from North.  The total flux density of the disc at this wavelength is set by $F_\nu$. Since the stellar emission is not significantly detected (see figure \ref{fclean}) we extrapolate the PHOENIX model \citep{brott05} fit of \citet{matthews14} to give a stellar flux density at this wavelength of $F_*=$17~$\mu$Jy. Although the star is not significantly detected, there is a clear peak near the phasecentre but offset by approximately $+0.4\arcsec$ in right ascension and $-0.2\arcsec$ in declination. This lies within the pointing accuracy of ALMA \footnote{$\sim$1.5{\arcsec} given the resolution and signal-to-noise ration of 1.1. See \url{https://help.almascience.org/index.php?/Knowledgebase/Article/View/319/0/what-is-the-astrometric-accuracy-of-alma} for the equation.} and so we assume that this is the location of the star. For the radial temperature profile we treat the grains as blackbodies.

\begin{figure*}
	\centering
	\includegraphics[height=4.7cm]{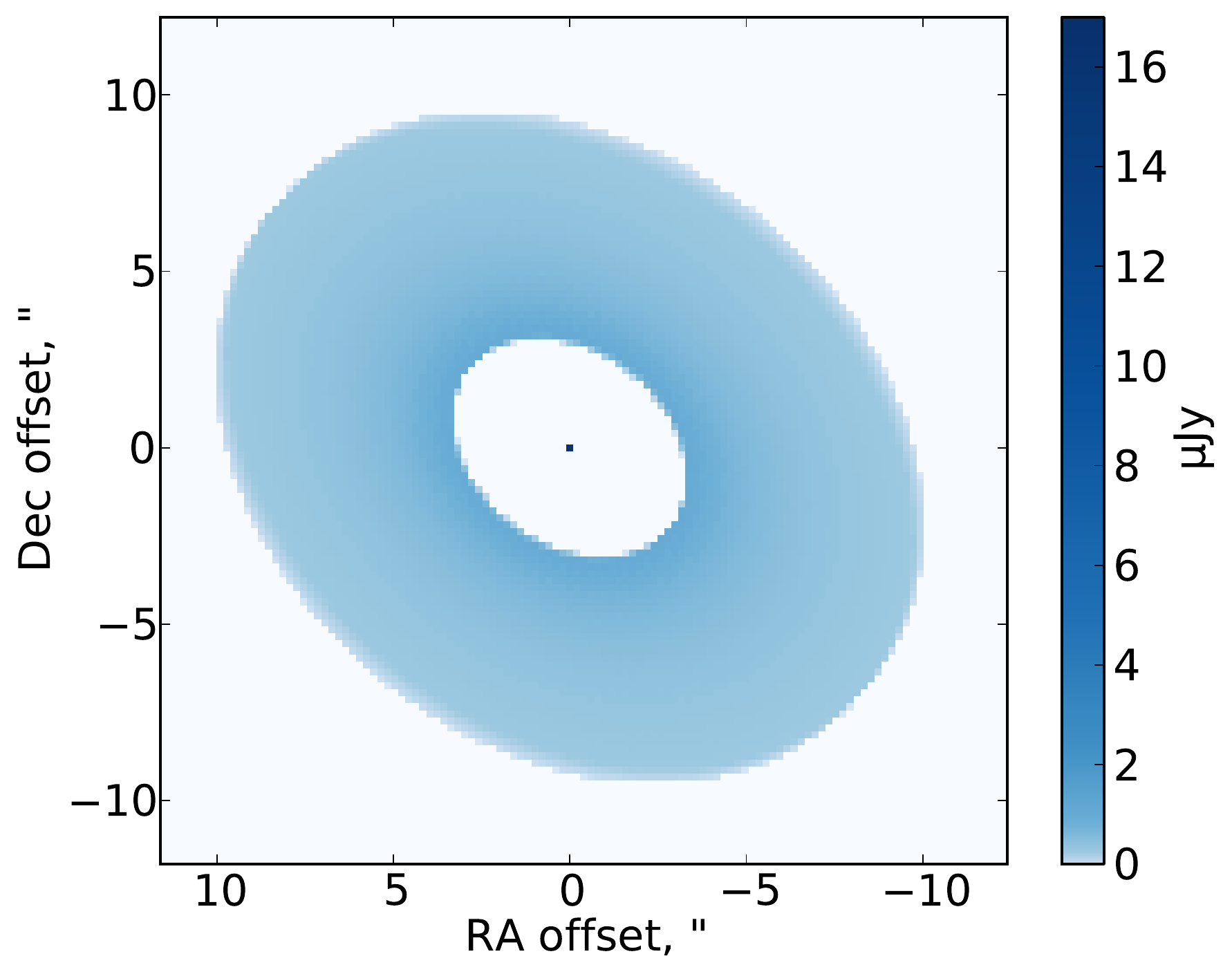}
	\includegraphics[height=4.7cm]{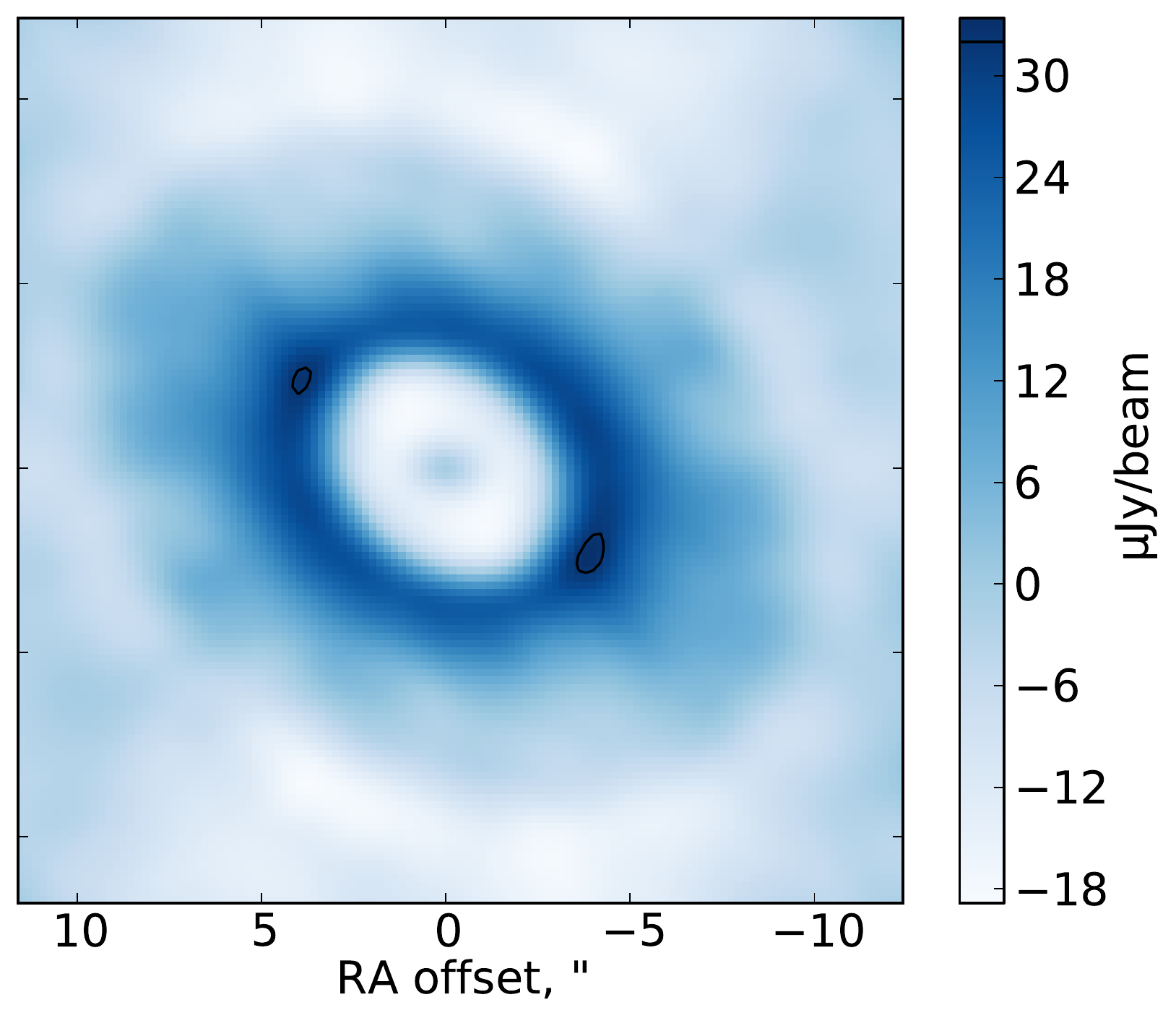}
	\includegraphics[height=4.7cm]{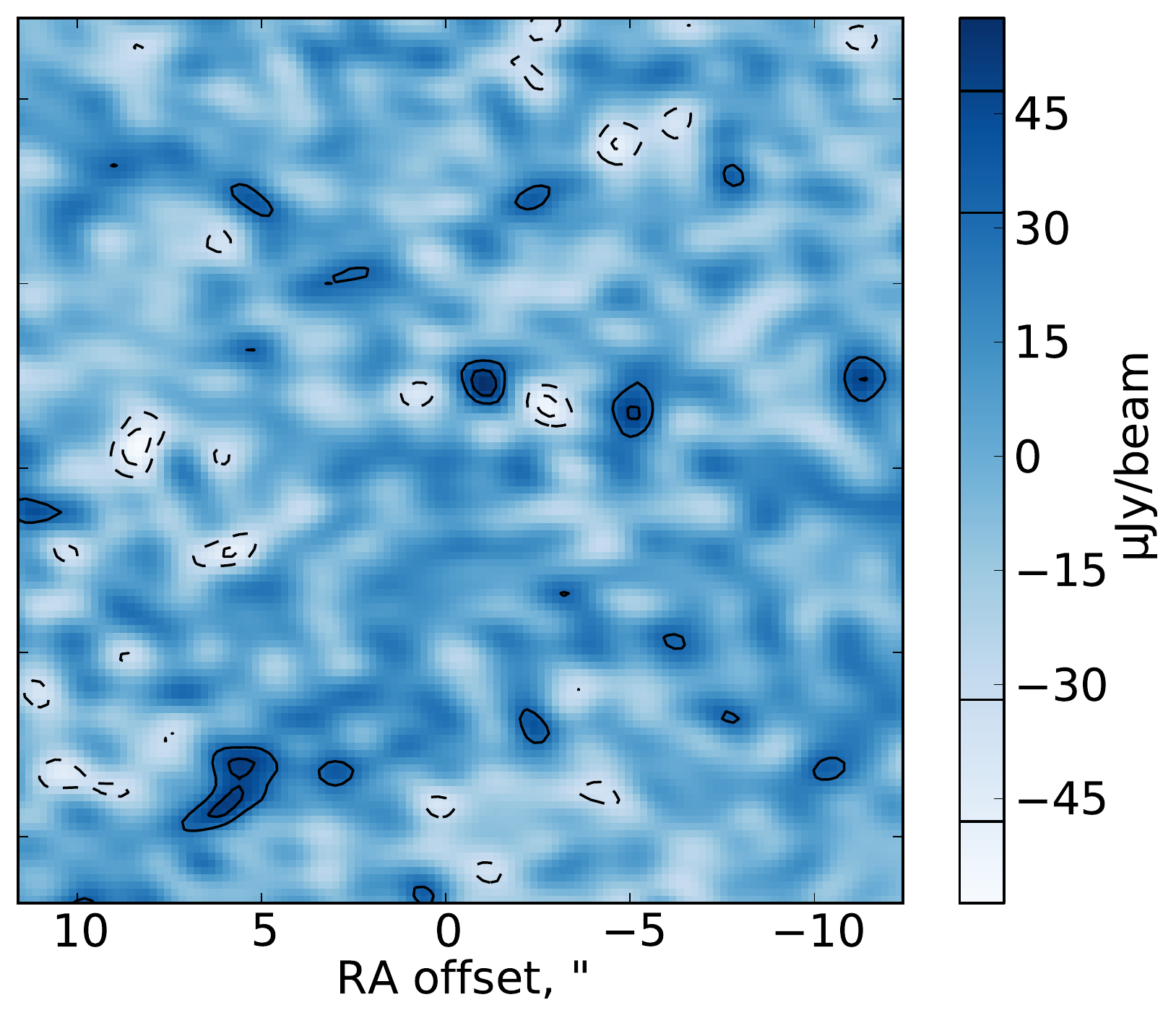}
	\caption{The modelling process is conducted by taking the sky plane image of the model (left), multiplying by the primary beam and convolving with the dirty beam to produce a convolved image (middle), then subtracting this from the dirty image to produce a map of residuals (right). The images shown here are for the best fitting model and contours are at significance levels of $\pm2$ and 3$\sigma$.}
	\label{fresid}
\end{figure*}

\begin{figure}
	\centering
	\includegraphics[width=0.45\textwidth]{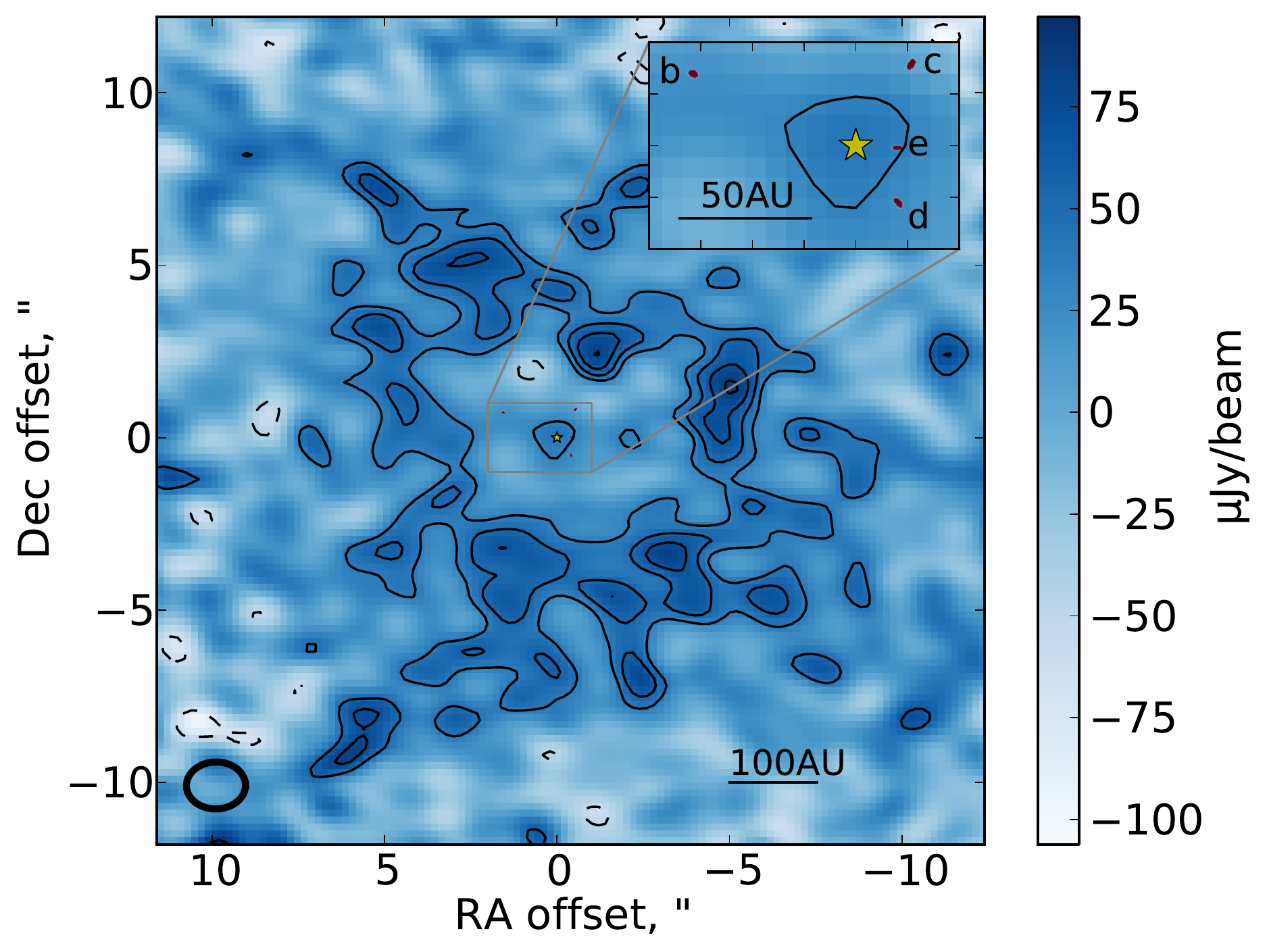}
	\caption{Restored image using the best fitting model (i.e the primary beam attenuated model convolved with a beam of size 1.7\arcsec$\times$1.3{\arcsec} -- shown by the black ellipse -- added to the residuals and primary beam corrected). Contours start at $2\sigma$ and increase in increments of $1\sigma$. This also shows the planets as seen in the K2 band with SPHERE-IRDIS on the VLT \citep{zurlo16}.}
	\label{frestore}
\end{figure}

To accurately replicate the observation process, the model image is first attenuated by the primary beam -- particularly important for this disc as the outer edge is potentially a significant fraction of the primary beam size and so we are likely to lose flux far from the phase centre. It is then convolved with the dirty beam (the Fourier transform of the sampling function) and the likelihood ($-\chi^2/2$) is then calculated by comparison with the dirty image. The high resolution of the image compared to the beam size has the side effect of introducing correlated noise. To account for this, the value of $\sigma$ used in the $\chi^2$ is modified by a noise correlation ratio given by the square root of the beam area in pixels. The likelihood is then passed to \texttt{emcee}, which is run with 120 walkers and for 2500 timesteps. Uninformative priors have been assumed for all the parameters, i.e. they are uniform in $R_{\rm{in}}$, $R_{\rm{out}}$, $\gamma$, $\ln(F_{\nu})$, $\cos(I)$ and $\Omega$.
The best fit parameters and their uncertainties are shown in table \ref{tres}. All parameters are fit to an accuracy or around 10\%. This may seem surprising for the outer edge as it extends to a region where the signal-to-noise ratio is low, but the strong interaction between the parameters helps constrain the outer edge as strongly as the other parameters. The best fit model, convolved model and residuals are shown in figure \ref{fresid}. Three residuals at $\sim$3$\sigma$ are seen in or near the disc at locations relative to the star of (5.7,-8.2), (-0.9,2.2) and (-4.9,1.3). Deeper observations are necessary to determine whether these are real and comoving with the star. The residuals are otherwise smooth and the RMS in the residual image is equivalent to the noise measured in section \ref{sobs}. A restored image of the disc is shown in figure \ref{frestore} showing how the disc should look to a single dish telescope i.e. removing the artefacts of the Fourier transform. 
\setlength{\tabcolsep}{5pt}
\begin{table}%
\begin{tabular}{cccccc}
$R_{\rm{in}}$ (AU)&$R_{\rm{out}}$ (AU)&$\gamma$&$F_{\nu}$ (mJy)&$I$ ($^\circ$)&$\Omega$ ($^\circ$) \\
\hline
$145^{+12}_{-12}$&$429^{+37}_{-32}$&$-1.0^{+0.4}_{-0.4}$&$2.8^{+0.5}_{-0.4}$&$40^{+5}_{-6}$&$51^{+8}_{-8}$ \\
\end{tabular}
\caption{Best fit parameters and uncertainties from the MCMC modelling (see section \protect\ref{smod}). The uncertainties given are the 16th and 84th percentiles.}
\label{tres}
\end{table}

\section{Discussion}
\label{sdisc}
\subsection{Comparison with previous observations}
The modelling of far-IR data by \citet{matthews14} predicts a planetesimal belt between roughly 100 and 310~AU whereas prior sub-mm interferometric observations using the SMA by \citet{hughes11} found an inner edge at $\sim$150~AU. Our inner edge of $145^{+12}_{-12}$ clearly reinforces the SMA result and is discrepant with the far-IR modelling. This discrepancy in inner edge at different wavelengths could potentially be real. The shorter wavelength observations are more sensitive to smaller grains. These grains are potentially susceptible to Poynting-Robertson drag, which will cause them to spiral in towards the star from the planetesimal belt as is seen in other discs such as Vega \citep{muller10}. The discrepancy in outer edge is significant and is likely due to the assumption by \citet{matthews14} of modelling the planetesimal belt and outer halo as two distinct components. In reality, there is likely some overlap between these components and the ALMA observations are better representing the actual outer edge of the planetesimal belt since the smaller grains in the halo are too faint at mm wavelengths to contribute significantly to the emission.

\citet{matthews14} find the disc to be oriented such that $I=26\pm3\degr$ and $\Omega=64\pm3\degr$ as opposed to our values of $I=40^{+5}_{-6}$ and $\Omega=51^{+8}_{-8}$. Whilst these parameters are consistent at the 95\% level, there is a possibility that there is a real difference between the orientation of the planetesimal belt and the small grain population similar to the warp seen in the $\beta$ Pic disc \citep[see][and references therein]{millar15}. Alternatively, it could be that emission from the background cloud (see section \ref{sobs}) or asymmetries in the disc (that have been assumed to be insignificant) are biasing the model fits in different ways. Given our knowledge of planetary system formation and evidence from our own Solar System \citep{brown04a}, it is expected that most planetary systems will be roughly coplanar unless they have undergone an instability. The latest astrometric fit to the observations of the planets from \citet{zurlo16} finds an inclination of 30$\pm5\degr$. A large difference in inclination could potentially tell us something about the dynamical history of the system but a 10$\degr$ difference, especially given the uncertainties, is not much larger than the inclination variations in the Solar System's planets and disc \citep[around $2\degr$,][]{brown04a}. 

An extrapolation of the total disc flux density based on the SCUBA and Herschel flux densities \citep{matthews14} predicts a value of 3.5~mJy at 1.3~mm. At $F_{1.3mm}=2.8^{+0.5}_{-0.4}$~mJy, the result found here is somewhat lower than the extrapolated value and may mean that the sub-mm slope is slightly steeper than previously thought.

\subsection{Interaction between planets and the disc}

The ability of planets to open gaps in circumstellar disks and to carve the inner edges of outer debris belts has been widely studied both analytically and numerically. This phenomenon is caused by the chaotic zone a planet possesses and within which mean motion resonances overlap. Therefore particles in the chaotic zone are unstable and will be removed on short dynamical timescales. The width of a planet's chaotic zone mainly depends on the mass and location of the planet \citep{wisdom80,duncan89}, but also on its eccentricity \citep{pearce14}. Consequently, it is possible to link the location of a debris belt's inner edge to the mass and orbital characteristics of the belt shaping planet. 

In the case of the HR 8799 system, it is expected that planet b (the outermost planet) is responsible for shaping the debris ring inner edge. The latest astrometric fit to all the observations of the planets is given in \citet{zurlo16}. Based on their orbital fit and using equations (9) and (10) of \citet{pearce14}, which gives the highest estimate of the width of the chaotic zone of all the methods mentioned previously, we expect an inner edge between 100-110 AU (for masses between 4-9\,M$_J$). An alternative method for finding the orbits of the planets is to run simulations of the dynamical stability of the planetary system as done by \citet{gozdziewski14}. They find that the planets must be in Laplace resonance to keep such a packed system of very massive planets stable. In contrast with the fit of \citet{zurlo16}, who assumed planets on circular and non coplanar orbits, their models involve planets on coplanar and eccentric orbits, which will produce a more distant inner edge for the disc. Although the nominal eccentricity of planet b in their fit creates an inner edge further out, it is not predicted to go beyond 106-116 AU with the largest estimates.
  
In practice, the inner edge of the disc can be further out than this as the analytical equations assume a fixed orbit whereas perturbations from the other planets on planet b can vary its orbit, thereby varying the chaotic zone. If planet b can reach an eccentricity of at least 0.3 (higher if the planet mass is less than 9~M$_J$) for a prolonged period of time, then it could clear planetesimals out to 145~AU. In order to capture these effects, n-body simulations are necessary. \citet{moro10} ran such simulations and they did find that it is possible to have an inner edge as far out as 150~AU. This may be due to variations in the orbital eccentricity of planet b coupled with the effect of the Laplace resonance, however, it may result from the low resolution of their simulations (one particle every 0.5 AU).  

All this opens up the possibility for another planet beyond the orbit of planet b. Such a planet would need to be beyond the chaotic zone of planet b to have any possibility of being stable. Using the equations of \citet{pearce14}, a planet at the edge of planet b's chaotic zone (110~AU) on a circular orbit would need to be $\sim$1.25~M$_J$ to create an inner edge at 145 AU. If it was further out or on an eccentric orbit, the required mass of the extra planet would be lower. This is lower than the limits from current observations, which can only reach multiple Jupiter masses \citep{zurlo16}.

The modelling in this paper assumes that the disc has sharp edges and is described by a single power law. In reality there are likely particles within the inner edge and the radial distribution may be better described by a rising then falling power law or a Gaussian-like distribution. The low level of signal-to-noise in the data does not warrant exploration of more detailed radial distributions, but they would be unlikely to affect our conclusions as the clearing of the chaotic zone is not 100\% efficient and so the inner edge referred to in the analytical and dynamical models is also not a sharp edge.

\section{Conclusions}
\label{scon}
This paper presents the ALMA observations of the debris disc around the star HR~8799. The observations were conducted in band 6 using a compact configuration. Using MCMC techniques, a parametric disc model is fit to the data. Assuming a single power law radial distribution, the inner edge is found to be $145^{+12}_{-12}$~AU and the outer edge is $429^{+37}_{-32}$~AU, whilst the inclination is found to be $40^{+5}_{-6}{^\circ}$ at a position angle of $51^{+8}_{-8}{^\circ}$. The inner edge found here agrees with that found by previous modelling of sub-mm data \citep{hughes11} but not with the models of the far-infrared data \citep{su09,matthews14}. It is also inconsistent with the edge of planet b's chaotic zone given its current orbit. This likely means that either planet b's orbit has varied over time or a smaller planet exists beyond its orbit. The outer edge is clearly much further out than the models that fit to the infrared data \citep{su09,matthews14}. This shows the benefit of sub-mm observations, which only detect dust in the planetesimal belt, whereas with infrared observations it is difficult to distinguish between the planetesimal belt and the halo of small grains being pushed out by radiation pressure. Finally, the CO detected in these observations confirms the prior observations that the CO in the cloud has the same radial velocity as that of the star and the clump in CO is found to overlap with the disc.

\section*{Acknowledgements}
We thank the referee for useful suggestions that helped improve the manuscript. The authors thank Alice Zurlo for providing the SPHERE data, Ed Fomalont for helpful discussions and Grant Kennedy for providing the stellar fit. M.B. and V.F. acknowledge support from FONDECYT Postdoctral Fellowships, project nos. 3140479 and 3150106. M.B., A.J., S.C. and J.C. acknowledge financial support from the Millennium Nucleus RC130007 (Chilean Ministry of Economy). A.J., S.C. and J.C. acknowledge support from FONDECYT projects
1130857, 1130949 and 1141175. A.J. and J.C. acknowledge support from BASAL CATA PFB-06. A.J. acknowledges support by the Ministry for the Economy, Development, and Tourism's Programa Iniciativa Cient\'{i}fica Milenio through grant IC\,120009, awarded to the Millennium Institute of Astrophysics (MAS). L.M. acknowledges support by STFC and ESO through graduate studentships and by the European Union through ERC grant number 279973. R.B. is supported by CONICYT-PCHA/Doctorado Nacional. This paper makes use of the following ALMA data: ADS/JAO.ALMA\#2012.1.00482.S. ALMA is a partnership of ESO (representing its member states), NSF (USA) and NINS (Japan), together with NRC (Canada) and NSC and ASIAA (Taiwan), in cooperation with the Republic of Chile. The Joint ALMA Observatory is operated by ESO, AUI/NRAO and NAOJ. The National Radio Astronomy Observatory is a facility of the National Science Foundation operated under cooperative agreement by Associated Universities, Inc. 
This research made use of Astropy, a community-developed core Python package for Astronomy \citep{astropy13}.

\bibliographystyle{mn2efix}
\bibliography{thesis}{}

\bsp
\end{document}